\documentclass[conference]{IEEEtran}
\pdfoutput=1 

\usepackage[pdflang={en-US},pdftex]{hyperref}
\usepackage[]{bookmark}
\usepackage{balance}       
\usepackage{graphics}      
\usepackage[T1]{fontenc}   
\usepackage{txfonts}
\usepackage{mathptmx}

\usepackage{color}
\usepackage{booktabs}
\usepackage{textcomp}

\usepackage{microtype}        
\usepackage{ccicons}          

\usepackage{todonotes}
\usepackage{float}

\def\plaintitle{Virtual Breathalyzer}

\def\emptyauthor{}
\def\plainkeywords{Machine Learning; Internet of Wearable Things; Intoxication Detection; Smart Watch; Google Glass; Sensors }

\makeatletter
\def\url@leostyle{%
  \@ifundefined{selectfont}{
    \def\UrlFont{\sf}
  }{
    \def\UrlFont{\small\bf\ttfamily}
  }}
\makeatother
\def\pprw{8.5in}
\def\pprh{11in}

\definecolor{linkColor}{RGB}{6,125,233}
\hypersetup{%
  pdftitle={\plaintitle},
  pdfauthor={\emptyauthor},
  pdfkeywords={\plainkeywords},
  pdfdisplaydoctitle=true, 
  bookmarksnumbered,
  pdfstartview={FitH},
  colorlinks,
  citecolor=black,
  filecolor=black,
  linkcolor=black,
  urlcolor=linkColor,
  breaklinks=true,
  hypertexnames=false
}

\begin{document}

\title{\plaintitle}

\author{\IEEEauthorblockN{Ben Nassi , Lior Rokach, Yuval Elovici}
	\IEEEauthorblockA{	Dept. of Software and Information Systems Engineering\\Ben-Gurion University of the Negev, Israel\\
		\{nassib,liorrk,elovici\}@post.bgu.ac.il}
}

\maketitle
\begin{abstract}
  Driving under the influence of alcohol is a widespread phenomenon in the US where it is considered a major cause of fatal accidents. In this research we present a novel approach and concept for detecting intoxication from motion differences obtained by the sensors of wearable devices. We formalize the problem of drunkenness detection as a supervised machine learning task, both as a binary classification problem (drunk or sober) and a regression problem (the breath alcohol content level).  
  In order to test our approach, we collected data from 30 different subjects (patrons at three bars) using Google Glass and the LG G-watch, Microsoft Band, and Samsung Galaxy S4. We validated our results against an admissible breathalyzer used by the police. 
  A system based on this concept, successfully detected intoxication and achieved the following results: 0.95 AUC and 0.05 FPR, given a fixed TPR of 1.0. Applications based on our system can be used to analyze the free gait of drinkers when they walk from the car to the bar and vice-versa, in order to alert people, or even a connected car and prevent people from driving under the influence of alcohol.
\end{abstract}

\section{Introduction}
Every 51 minutes a person died in a motor vehicle accident caused by an alcohol impaired driver in 2013, a tragic statistic that represents more than 30\% of all US traffic-related deaths that year \cite{CDC}. The high rate of fatal accidents resulting from "driving under the influence" (DUI) reflects the devastating effects of alcohol consumption on driving (e.g reduced coordination, difficulty steering, and reduced ability to maintain lane position and brake appropriately).
\par DUI is a widespread phenomenon in the US. Yet in 2012, only 1\% of the people that reported episodes of alcohol impaired driving were arrested. Many counter measures are deployed in order to reduce the influence of driving while intoxicated. The most well-known strategy employed to catch intoxicated drivers  is the breath alcohol concentration (BrAC) test that measures the weight of alcohol present within a certain volume of breath \cite{SelfTestforBreathAlcohol}. This test is conducted with the breathalyzer device \cite{Breathalyzer} which uses the driver's breath as specimen/sample. The breathalyzer uses infrared or electromechanical tests, and sometimes even a combination of both, to determine the BrAC within a short amount of time.
\par BrAC limits vary by country causing the definition of drunkenness differ around the world. Table~\ref{tab:table1} contains the four most common BrAC thresholds (220, 240, 250, 350 micro-grams of alcohol per one liter of breath) used around the world. Based on these standards, anyone with a breath alcohol concentration measured by a breathalyzer above the defined threshold for a given country is considered to be intoxicated. In the US, the threshold varies widely between each state.
\begin{table}
	\centering
	\begin{tabular}{l l}
		{\small\textit{BRAC Threshold}}
		& {\small \textit{Countries}} \\
		\midrule
		220 & Scotland, Finland, Hong Kong  \\
		240 & Israel \\
		250 & Greece, Spain, New Zealand \\
		350 & Singapore, UK, Wales, Trinidad and Tobago \\
	\end{tabular}
	\caption{BrAC threshold (micro-grams of alcohol per one liter of breath) permitted in each country}~\label{tab:table1}
\end{table}

In this paper, we suggest a new approach for the detection of intoxication based on wearable technology. It is a known fact that alcohol consumption causes changes in people's movements. We hypothesize that these changes can be measured using wearable device's sensors in order to detect drunkenness using a trained machine learning model.
The expected contributions of this research stem from the following points:

\begin{enumerate}
	\item We show that existing widely used wearable devices can be used to identify the physiological indicators that
	imply drunkenness (in terms of body movement) based on free gait instead of traditional ad-hoc sensors that focused on the breath.	
	\item We formalize the task of the detection of intoxication as a supervised machine learning task based on body movement measurements derived from wearable devices. We used an actual breathalyzer (as used by police departments) in order to label our data and train our models to evaluate our results. 
	\item We analyze and compare the usefulness of each type of sensor and device for detecting drunkenness (in terms of body movement) based on free gait. 
	
\end{enumerate}

\par A system based on our approach can prevent people from driving under the influence with an alert provided after detecting intoxication unobtrusively as a person leaves a bar (after spending some time drinking) and walks to their cars. In the era of Internet of Things and connected cars, it might even be used to prevent people from driving under the influence of alcohol by triggering a connected car to prevent the ignition of the car when the car owner is detected as drunk.
 
\section{Related Work}
Although wearable devices have already ubiquitous, there has been no scientific work conducted in the area of intoxication detection using wearable devices by analyzing the movement and gait of subjects. However, such analysis \cite{mantyjarvi2005identifying,gafurov2007gait,lu2014unobtrusive,aminian2014gait,murata2013wearable} has been used in other areas for various purposes and tasks. 
\subsection{Gait Analysis}
Gait analysis has been studied for many years, even before the era of wearables devices. Up to a decade ago, researchers were using ad-hoc sensors specially designed for research purposes. Mantyjarvi \textit{et al}. \cite{mantyjarvi2005identifying} analyzed data collected from warn accelerometer devices in order to identify subjects by their gait. Gafurov \textit{et al.} \cite{gafurov2007gait} used a worn accelerometer for authentication and identification based on the subjects' gait, while Lu \textit{et al.} \cite{lu2014unobtrusive} showed that authentication from gait is also possible from smartphone sensors. Aminian \textit{et al.} \cite{aminian2014gait} analyzed accelerometer and gyroscope measurements from ad-hoc sensors that they designed to be worn on a shoe in order to explore gait.
Xu \textit{et al.} \cite{xu2012smart} presented a novel system for gait analysis using smartphones and three sensors located within shoe insoles to provide remote analysis of the user's gait. 
\subsection{Wearable Device Products}
In 2013, Google introduced Google Glass , and in recent years other companies have introduced smart watches, wristbands, and fitness trackers. The prevalence of these devices means that practical research can be carried out on existing products. In the area of movement analysis and wearable devices and products, Thomaz\textit{ et al.} \cite{thomaz2015practical} used smartwatch motion sensors in order to detect eating instances. Ranjan \textit{et al.} \cite{ranjan2015object} analyzed smartwatch sensors during specific home-based activities (such as turning on a light switch) to identify subjects based on hand gestures. In the field of emotion detection, Hernandez \textit{et al.} \cite{hernandez2014senseglass} analyzed head movement from Google Glass motion sensors in order to detect stress, fear and calm. Hernandez \textit{et al.} \cite{hernandez2015biowatch} analyzed smartwatch motion sensors to estimate heart and breathing rates. Mazilu \textit{et al.} \cite{mazilu2015gait} analyzed wrist movement to detect gait freezing in Parkinson's disease using data sensors of smart watches and wristbands. 
Gabus \textit{et al.} \cite{gabuskostopoulos2015increased} and \textit{Casilari et al.} \cite{casilari2015automatic} used a smartwatch in order to detect falls. 
Inou \textit{et al.} \cite{inoue2015mobile} used three accelerometers located on the right wrist, breast, and back hip in order to recognize nursing activities of nurses in a hospital.

\subsection{Intoxication Detection}
Despite of the volumes of related work, there has been a limited amount of research that addresses the domain of intoxication detection using the ubiquitous wearable technology. In the area of movement analytics as a method to detect drunkenness, there are a few works that relied on the fact that the short-term effects of alcohol on subjects cause impairment to movement, gait, and balance.

Kao \textit{et al.} \cite{kao2012phone}, analyzed the accelerometer data collected from the smartphones of three subjects and compared the step times and gait stretch of sober and drunk gaits. This research was limited in scope in that it only used smartphone devices and three subjects. Also, it wasn't aimed at detecting whether a person was drunk given data collected from the device; instead the study compared differences in drunk and sober gait.  
Arnold \textit{et al.} \cite{arnold2015smartphone} investigated whether a smartphone user's alcohol intoxication level (how many drinks they had) can be inferred from their gait. They used time and frequency domain features extracted from the device's accelerometer to classify the number of drinks a subject consumed based on the following ranges: 0-2 drinks (sober), 3-6 drinks (tipsy), or 6+ drinks (drunk). Their results were not validated against a real breathalyzer, and the data was only collected from a mobile phone.
Several works have utilized ubiquitous technology to detect drunkenness based on driving patterns. Dai \textit{et al.} \cite{dai2010mobile} and Goswami \textit{et al.} \cite{goswami2014android} used mobile phone sensors and pattern recognition techniques to classify drunk drivers based on driving patterns. Other studies tried to detect drunkenness using different approaches. Thien \textit{et al.} \cite{thienhorizontal} and Wilson \textit{et al.} \cite{wilsonimage} attempted to simulate the HGN (Horizontal Gaze Nystagmus) test \cite{field-sobriety-test-review} in order to detect drunkenness using a camera (i.e., smartphone camera) and computer vision methods. Hossain \textit{et al.} \cite{hossain2016inferring} used machine learning algorithms to identify tweets sent under the influence of alcohol (based on text).

\section{Detailed Description of the Research}

Our study utilizes machine learning to understand changes in patterns of movement (movement analytics) that imply intoxication. We use wearable devices’ sensors to measure the effects of drunkenness on parts of the body. In this section we describe: (A) the factors to consider with regard to device selection, (B) the system we developed for the experiment and (C) our approach for detecting intoxication from wearable devices. 

\subsection{Selecting the Wearable Devices}

\begin{table}[]
	\centering
	\resizebox{\columnwidth}{!}{%
		\begin{tabular}{cccccc}
			\begin{tabular}[c]{@{}l@{}}			Criteria/\\Device \end{tabular} & Galaxy S4 & LG-G Watch & Google Glass & Microsoft Band  \\\hline
			Type 
			& Smart Phone
			& Smart watch
			& Smart Glass
			& Fitness Tracker\\\hline
			Sensors 
			& \begin{tabular}[c]{@{}l@{}}Accelerometer \\ Linear \\ Gyroscope \\ Gravity \\Compass\end{tabular} 
			& \begin{tabular}[c]{@{}l@{}}Accelerometer \\ Linear \\ Gyroscope \\ Gravity \\Compass\end{tabular}
			& \begin{tabular}[c]{@{}l@{}}Accelerometer \\ Linear\\ Gyroscope \\ Gravity \\Compass\end{tabular} 
			& \begin{tabular}[c]{@{}l@{}} Accelerometer\\ Gyroscope \end{tabular}\\\hline
			\begin{tabular}[c]{@{}l@{}}Maximum \\ Sampling \\ Rate\end{tabular}  
			& 180Hz
			& 200Hz
			& 100Hz
			& 62Hz\\\hline			
			\begin{tabular}[c]{@{}l@{}}			Software\\ Developer\\ Kit \end{tabular}
			& Android 
			& Android Wear
			& Android Wear 
			& Android \\\hline
			Connectivity 
			&\begin{tabular}[c]{@{}l@{}}			GSM\\WiFi\\ Bluetooth  \end{tabular}			  
			& \begin{tabular}[c]{@{}l@{}}			WiFi\\ Bluetooth  \end{tabular}
			& \begin{tabular}[c]{@{}l@{}}			WiFi\\ Bluetooth  \end{tabular}
			& Bluetooth \\\hline

		\end{tabular}
	}
	\caption{Sensors available in each device.}~\label{tab:table2}
\end{table}

\begin{figure*}
	\centering
	\includegraphics[width=1.75\columnwidth]{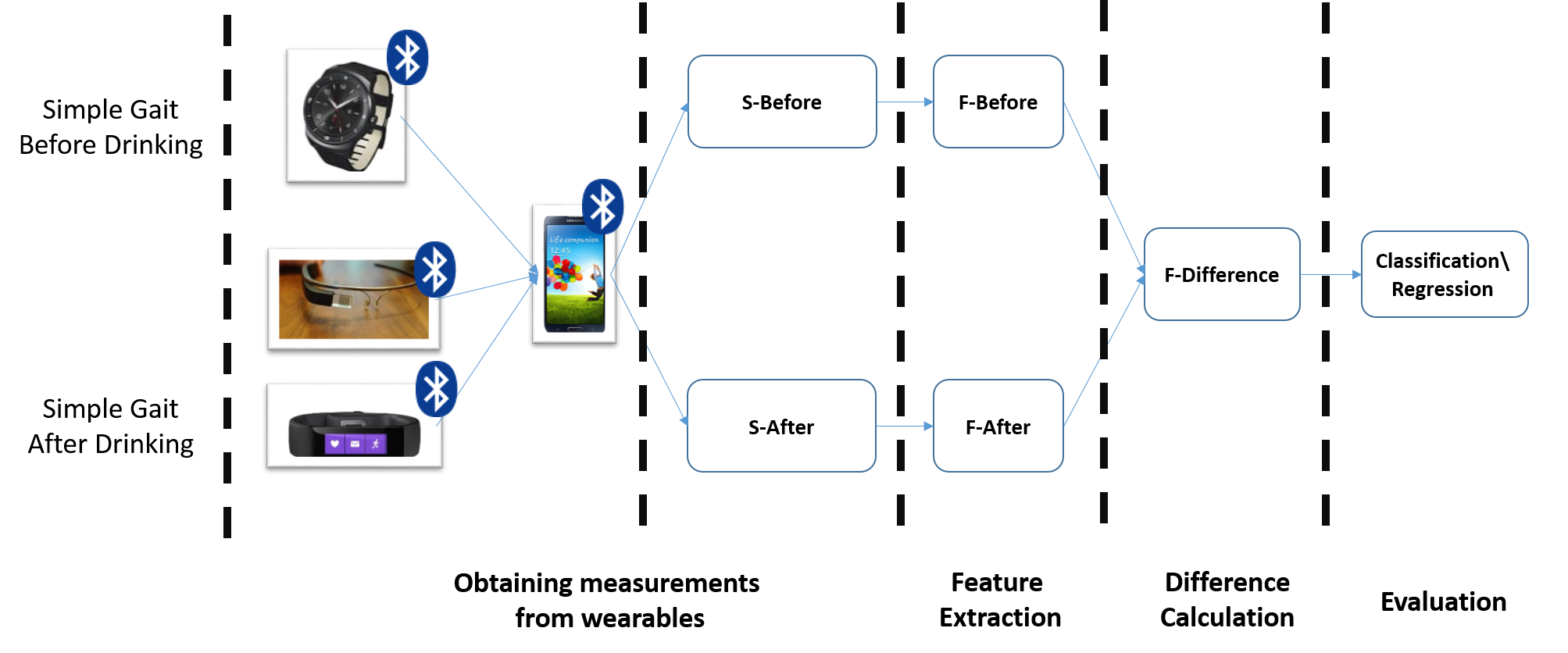}
	\caption{Overview of the system and our approach. Data is collected before and after drinking from the sensors of the devices and sent to the mobile phone using Bluetooth (S-Before and S-After measurements). The signals are filtered with SMA (simple moving average), and 2 sets of features are extracted (F-Before and F-After) from the signals. A calculation of the difference of the features is applied to extract a new set of features (F-Difference). The classification/regression model processes the instance using a learned model and outputs the result.}~\label{fig:figure1}
\end{figure*}

In order to  determine whether intoxication can be detected from changes in motion using wearable devices, we first have to select the devices to use in our experiment. There is a wide range of wearable devices available, and in order to focus our research efforts on a few specific devices, we defined a list of technical requirements for the devices to be used in our study.
\begin{enumerate}
\item Applicability - We wanted our research to be as applicable as possible for future use (i.e., application-oriented in order to develop a product based on our results). For this reason, we made the decision to utilize the sensors of existing products as opposed to specifically crafted sensors.
\item Built-in motion sensors - Since our research is based on detecting the effects of drunkenness on subjects' movements, the wearable device must include motion sensors.
\item Programmability - The wearable device must provide a software developer kit (SDK) which will allow us to develop the application that will sample the motion sensors.
\item Undependability and Connectivity - The device has the ability to be used and evaluated separately and as part of connected network. 
\item Versatility - Since we didn't know in advance which part of the body would be the most effective and accurate indicator of drunkenness, we wanted to work with devices that measure the movements of different parts of the body.
\end{enumerate}

The following devices were selected in accordance with the aforementioned specifications: Google Glass, the LG G-watch,  the Microsoft Band, and the Samsung Galaxy smartphone. 
\par Smartphones are very common and provide us an excellent infrastructure for our research. Regarding smartwatches and fitness trackers, according to a survey conducted by Global-Web Index with 170K participants around the global in 2014,  one out of six people already owns either smartwatch or fitness tracker \cite{GWI-Device-Summary}; they predict that smartwatches will make up over 50\% of the wearable market by 2020 \cite{smartwatch-report}.
Although the market share of Google Glass is low (only ~800K units were sold in 2014) \cite{BI-GoogleGlass-marketshare}, Google Glass was chosen because: (1) Business Insider expects unit sales of Glass to climb sharply in the next few years, to 21 million units in annual sales by the end of 2018\cite{BI-GoogleGlass-forecasts}, and (2) there is a large increase in smart devices for the head, such as VR glasses, AR helmets, and smart helmets for motorcycles. We chose Google Glass to represent the set of smart devices targeted for the head, a type of device that we believe will become common place in the next years.

\par Table~\ref{tab:table2} presents the motion sensors, SDK and connectivity of each of the selected devices. The devices selected provide the ability to measure movement in four parts of the body: the right hand, left hand, head, and rump.

\subsection{Developing the system for the experiment}

\par In order to conduct an experiment to collect data, we used the SDKs to develop a client application for each of the devices. The client runs as a background service and waits for start/stop commands in order to start/stop the recording of all of its available motion sensors (Table~\ref{tab:table2}). We paired each of the wearable devices with the mobile device using Bluetooth communication and added the functionality of sending start/stop commands to each of the clients from the mobile device. The data was sampled with the maximum rate supported by each device and was recorded as a time series in nanoseconds. When the stop command was received from the mobile phone, each of the clients sent the data it sampled to the mobile phone using Bluetooth. 
\begin{figure*}
	\centering
	\includegraphics[width=2.0\columnwidth]{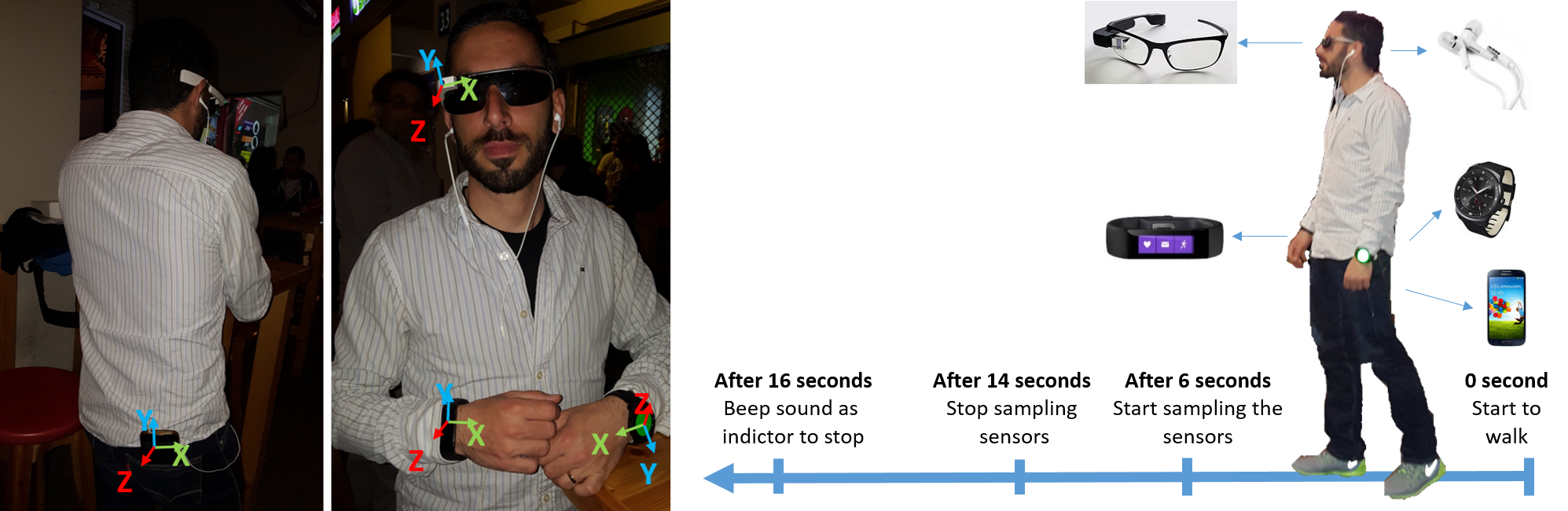}
	\caption{From left to right: (a) The x, y, and z axes of the four devices. (b) The experiment stages. }~\label{fig:figure2}
\end{figure*}
\subsection{Our Approach}
We propose a new approach to detect intoxication (using the platform described in the previous section) based on changes in movements extracted from two measurements of simple gait at two times (associated with the two states of the subject): the first movement sample/measurement is taken before the subject has begun to drink alcohol, and the second sample is taken after the subject has finished drinking (our experiment is described in greater details in the next section). 
\par A subject's movement sample/measurement consists mainly of a time-series of values obtained from each of the wearable devices used. Since each wearable device contains 4-5 motion sensors, each sample consists of almost 20 streams (time-series) of a sensor's data.
\par Given person p and his/her weo samples: s-before (measurement taken before alcohol consumption) and s-after (measurement taken after alcohol consumption), we process them as follows:
\begin{enumerate}
	 \item Feature Extraction - 
	 We extract two sets of features: the f-before set (extracted from s-before) and the f-after set (extracted from s-after). 	 
	 \item Difference Calculation - 
	 We calculate a new set of features called the f-difference. These features represent the difference (for each feature) between the f-after and f-before. The difference signifies the effects of the alcohol consumption on the subject's movement. 	 
	\item Labeling - 
	We label the sample of each subject with the result of a professional breathalyzer (one that is used by law enforcement) taken immediately after s-after is measured.  
	\item Evaluation -
	 We apply supervised machine learning algorithms on a new set of data, consisting of the union of the f-differences calculated for each of the subjects and calculate the error. 
	
\end{enumerate}

By processing the samples following the four steps described above, we are able to identify the physiological indicators that imply drunkenness (in terms of body movement) based on free gait. This approach can be used in order to prevent people from driving under the influence. By obtaining two samples of an individual's gait (the first sample obtained on the subject's way from their car to a bar, and the second taken as they return to their car from the bar).
Figure~\ref{fig:figure1} provides an overview of our approach.
\section{Experimental settings}
\begin{figure}
	\includegraphics[width=1.0\columnwidth]{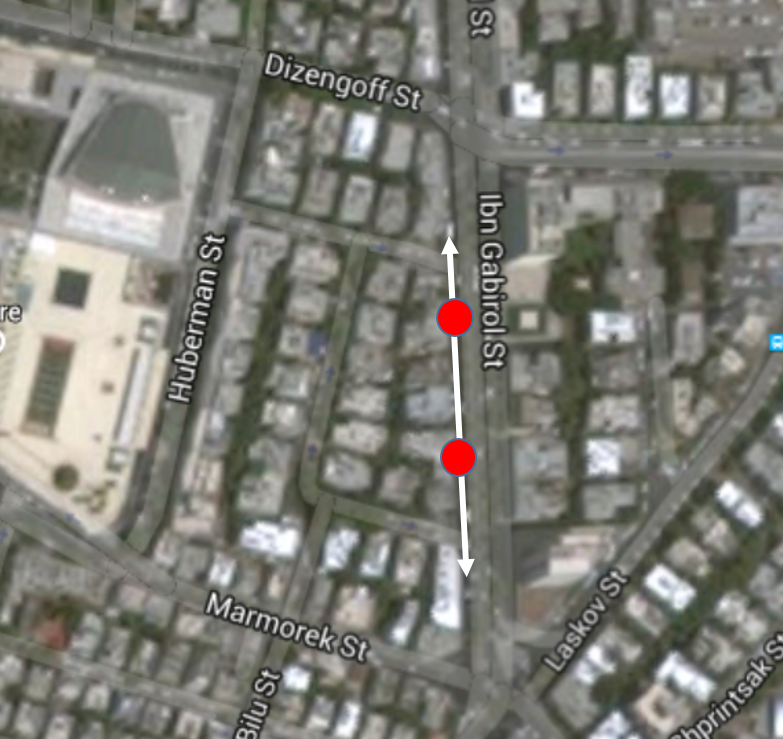}
	\caption{Aerial view of the street where the experiment took a place. The 2 red dots mark 2 out of the 3 bars and white line shows the path that our subjects walked during the experiment.}~\label{fig:figure660}
\end{figure}
\begin{figure*}
	\centering
	\includegraphics[width=2.0\columnwidth]{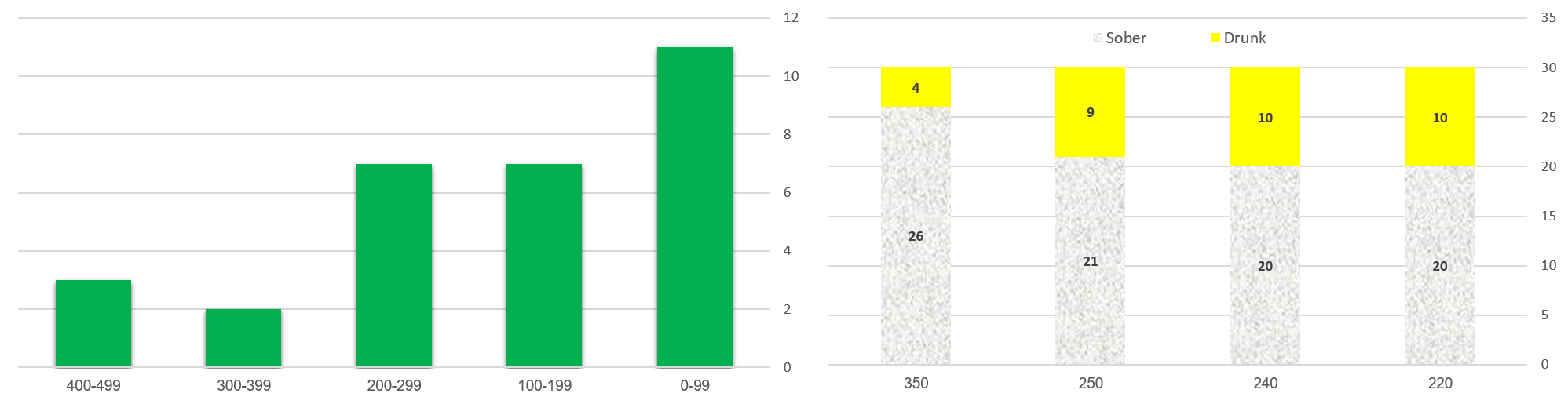}
	\caption{From left to right. (a) Breathalyzer samples - The bars represent the results of the subjects' breathalyzer tests (the amount of micro-grams of alcohol per one liter of breath). (b) A breakdown of the subjects’ state (sober/drunk) at various BAC levels.}~\label{fig:figure2000}
\end{figure*}
Our subjects were outfitted with the devices as follows. They each wore Google Glass on their head, an LG G watch on their left hand, and a Microsoft Band on their right hand (see Figure~\ref{fig:figure2}). The subjects who had a right rear pocket also carried a mobile device there. Those subjects without a right rear pocket were only outfitted with three devices (the mobile was not part of their experiment). Each subject also wore headphones. 

\subsection{The Experiment}

The process of obtaining gait samples included the following steps. First, each subject was instructed to walk with the devices and headphones until he/she heard a beep in their headphones (for a total of 16 seconds). During the subject's gait, the devices' clients sampled the sensors for eight seconds, a time period that started on the sixth second of the experiment and continued until the fourteenth second. The stages of the experiment are presented in Figure~\ref{fig:figure2}. 
\par We decided that using 16 seconds of free gait is the optimal way to conduct the experiment and obtain the samples for the following reasons: (1) The gait is probably the best way to ensure that the devices are carried with the person instead of standing on a table during a conversation. (2) Because we can automatically turn the model on when the GPS in a device sense a movement, instead of relying on the subject to remember to turn on the application. (3) The model will be relevant, even for people whose cars are parked very close to the bar. (4) Using free gait instead of a specific challenge for our model will ensure that potential subjects won’t be put off by the nature of the challenge and allow the model to be used in real life situations.

\par The experiment was conducted in two sessions. The first session took place before the subjects had their first drink. The second session took place 15 minutes following the subjects' last drink, just before they intended to leave the bar. It was crucial to wait 15 minutes after the last drink, because in addition to the gait samples, our subjects provided us with breath samples using our Drager Alcotest 5510 breathalyzer, and 15 minutes is the same amount of time that police wait before obtaining a suspect's breath sample. We used this type of breathalyzer, because it is a professional breathalyzer used by police departments in different countries around the world. This breathalyzer outputs results in micro-grams of alcohol per one liter of breath. We labeled each sample with the breathalyzer result (BrAC). 

\par In order to sample as many people as possible, our experiment took place at three different bars that offer an "all you can drink" option. We waited for people to arrive at the bar, and just before they ordered their first drink, we asked them to participate in our research (participation entailed providing a gait sample during two brief experimental sessions while wearing wearable devices, as well as providing a breath sample). 

\par We asked 30 different subjects to participate in our research. Each person was instructed to walk (while wearing the devices) in any direction they wished until they heard a beep in the headphones (this was done in order to simulate the path between the bar and the subject’s car). The subjects were outfitted with three to four devices(as mentioned previously, depending on whether they had a right rear pocket). Participants were paid for their participation in the study (each subject received the equivalent of 13 USD in local currency). Figure~\ref{fig:figure660} presents an aerial view of two of the three bars and the path that our subjects took during the experiment.

\subsection{Data Analysis} 

Our experiment is based on 60 samples taken from 30 individuals on five nights at three bars. Table~\ref{tab:table1111} provides information about subjects. Most of our participants were in their early 20's, which according to US National Highway Traffic Safety Administration (NHTSA) \cite{nhtsa} is the group considered to have the highest risk of causing fatal accidents due to alcohol consumption (in 30\% of the resulting from intoxicated drivers in 2014, the drivers were between the ages of 21 and 24). 

\par Figure~\ref{fig:figure2000} presents the analysis and distribution of the breathalyzer results. Most of our samples (89\%) are labeled with breathalyzer result (the amount of micrograms of alcohol per one litter of breath) in the 0-400 range. Our data needed to include samples of both sober and drunk states. This was crucial to the model creation phase (described later) in order to learn the movement differences that imply intoxication, as well as the differences that don't imply intoxication. The breakdown of the subjects {sober/drunk} states  is also presented in Figure~\ref{fig:figure2000} according to the thresholds presented on Table~\ref{tab:table1}. At the lower thresholds of alcohol concentration (220,240,250) the data is distributed, such that 27\%-35\% of the total number of subjects were considered to be intoxicated. At the highest threshold (350) 16\% of the subjects were considered drunk.
\begin{table}[h]
	\centering
	\resizebox{\columnwidth}{!}{%
		\begin{tabular}{l c c c c}
			{\small\textit{Gender}}
			& {\small \textit{\begin{tabular}[c]{@{}l@{}}Number of \\ Participants\end{tabular} }} 
			& {\small \textit{Age(Year)}} 
			& {\small \textit{Height(CM)}} 
			& {\small \textit{Mass(KG)}} \\
			Male   & 24 (80\%) & $24.1\pm3.6$ & $176.4\pm9.2$ & $73.1\pm10.5$ \\\hline
			Female & 6 (20\%)& $24.5\pm5.9$ & $168.5\pm4.5$ & $60\pm4.5$  \\\hline
		\end{tabular}
	}
	\caption{Details of the participants in the experiment. Each cell presents the average and the standard deviation}~\label{tab:table1111}
\end{table}

\subsection{Fusion of the Sensors}
\par Each subject contributed two breath and gait samples (obtained in two sessions - before and after drinking). Each gait sample is comprised of sensor readings (measurements) obtained from three or four different wearable devices (depending on whether the subject had a right rear pocket for a mobile phone). Each wearable device contained three to five different motion sensors (as described in Table 2) that sampled during the experiment. This amounts to measurements from 15-20 sensors  obtained in each session from the various devices. 
\par Each sensor sample produces a vector with four dimensions: time in nanoseconds (from the beginning of each experiment) and the values x, y, z that represent the values for each axis as a real number. The entire set of samples obtained during each session from a specific sensor completes the signal that is represented as a time series of values in each of the axes (x,y,z). Figure~\ref{fig:figure2} presents the axes of each device used in the experiment. 

\section{Feature Extraction}
Given person p with two measurements (s-before and s-after) consisting of the sensor's data, with four dimensions (time,x value,y value, and z value) for each of of the wearable-devices, we consider the time-series of each axis of each measurement as an independent signal (e.g, the x axis of the smart-watch accelerometer of measurement s-before).
Given two signals s-before and s-after (measured from the same person and the same axis) from the same wearable-device, we extract the correlative vectors f-before and f-after, the features of each signal, from four different types (the entire set of features is described on Table~\ref{tab:table3}):
\begin{enumerate}
	\item Fast Fourier Transform: FFT was used on each axis of the signals in order to transform the signal to the frequency domain and extracted the features described in previous works \cite{mantyjarvi2005identifying,hernandez2015biowatch}. Such features may indicate physiological changes resulting from alcohol consumption that are associated with frequency of movement (e.g., fewer steps per second as obtained from the mobile phone).
	\item Statistic features: Again, we extracted statistic features mentioned in previous works \cite{lu2014unobtrusive,thomaz2015practical,zhang2011feature,zhang2012motion}, because they may indicate physiological changes associated with intoxication, such as decreased average acceleration of hand movements measured by motion sensors on the smart watch or sharp head movement obtained by the maximum values of Google Glass' gyroscope. 
	\item Histogram features: We presented the signals as histograms as presented in previous studies \cite{basaran2014classifying,Hazan:2015:NRM:2825041.2825062}. We extracted histograms from each signal that represent the values (in integers) between the maximum and minimum value of each sensor on each axis. Since our data was represented as real numbers, we round the values of the signals, count them into the histogram, and normalize them by dividing the length of each signal. We extracted this set of features, because we thought that they would be helpful to us in detecting differences in the patterns of movement (and specifically, the distribution of the movement) resulting from alcohol consumption. 
	\item Known gait features:  This type of features has been shown to yield good results in previous studies \cite{zhang2011feature,zhang2012motion}

\end{enumerate}
We repeat the feature extraction process for each signal and concatenate the entire set of features extracted for the measurement of 1 large vector.
The f-before-complete and f-after-complete vectors are vectors that have the same length extracted from all of the signals of two different measurements (s-before and s-after). Each index in the vectors is associated with a feature from the features' list for a specific signal. The index i in vectors f-before-complete and f-after-complete represents the same feature being extracted from the s-before and s-after (e.g., the value of index one in vectors f-before-complete and f-after-complete is the mean acceleration measured from the accelerometer of Google Glass in axis x from s-before and s-after). 
We consider the entire set of vectors of features extracted from all of the signals of specific measurement (e.g., all the features' vectors being extracted from s-before) as one long vector
\par We calculate a new set of features called f-difference as follows: for each one of the features we extract (described in Table~\ref{tab:table3}), we calculate the difference between its value in f-before-complete and its value in f-after-complete.
\par
We label these differences with the breathalyzer results (to allow us to use supervised machine learning algorithms). We consider each f-difference vector labeled by a breathalyzer result as an instance. We consider the union of these 30 instances as our data-set, representing the indicators associated with subjects' level of intoxication (measured by the breathalyzer).

\begin{table}
	\centering
	\small
	\resizebox{\columnwidth}{!}{%
		\begin{tabular}{l l}
			\midrule
			\begin{tabular}[c]{@{}l@{}}Statistics \\ Features\end{tabular}  
			& 
			\begin{tabular}[c]{@{}l@{}}	 Mean, Variance, Covariance, Standard Deviation,\\ Skewness, Min, Max, Median, Range, Root mean square,\\ Zero crossing rate, Mean crossing rate  \\\end{tabular}\\  		
			\hline
			\begin{tabular}[c]{@{}l@{}}FFT \\ Features\end{tabular}  
			& 
			\begin{tabular}[c]{@{}l@{}}	Energy, Top 4 frequency values,\\ Frequency bin that contains the frequency \\
				with maximum magnitude value  
				\\\end{tabular} \\ 
			\hline
			\begin{tabular}[c]{@{}l@{}}Histogram  \\ Features\end{tabular}  
			& 
			\begin{tabular}[c]{@{}l@{}}	Normalized histogram of the\\ values (each value is rounded to the nearest integer)
				\\\end{tabular} \\ 
			\hline
			\begin{tabular}[c]{@{}l@{}}Known \\Gait\\ Features\end{tabular}  
			& 
			\begin{tabular}[c]{@{}l@{}}	Movement intensity, Normalized signal magnitude area,\\ 
				Eigenvalues of dominant directions, Eigenvalues of \\dominant directions, 
				Correlation between acceleration \\along gravity and heading directions, 
				Average velocity\\ along heading direction, Average velocity along gravity \\direction,
				Average rotation angles related to \\gravity direction, Dominant frequency,
				Energy, \\Average acceleration energy, Average rotation energy
				\\\end{tabular} \\ 
			\hline
		\end{tabular}
	}
	\caption{ Features types extracted.}~\label{tab:table3}
\end{table}

\section{Results}
The feature extraction process resulted in 30 labeled instances extracted from 30 users, representing the differences between the extracted features before and after drinking. We used this data to train supervised machine learning models using two approaches. 
\par The first approach is to analyze the data as a regression task in which we try to predict the breathalyzer result given the differences in the subject's gait features. We also aim to minimize the RMSE (root mean square error) and MAE (mean absolute error). We consider a model to provide better results compared others if its MAE and RMSE are lower.
\par The second approach is to analyze the data as a classification task, with a goal of determining whether a person is drunk or sober according to known BrAC thresholds as measured using a breathalyzer. More precisely, we aim to train a model that predicts whether a person is intoxicated or not using differences in the subject's gait features. We chose to classify our instances as one of four thresholds {220, 240, 250, 350} (presented in Table 1). We consider an instance labeled by a breathalyzer result (BrAC) to be sober if its value is less than the threshold, and drunk if its value exceeds the threshold. In this approach we try to maximize the AUC (area under the curve) and the model's accuracy. In this task, we consider a model to provide better results than others if it has higher AUC results.
\par In order to classify/predict an outcome of an instance given the values of the features, we used ensemble machine learning algorithms based on a decision tree. A decision tree recursively partitions the independent variables' space into subspaces such that each subspace constitutes the basis for a different prediction. A single decision tree usually has limited prediction performance. One way to improve the prediction performance is to build a forest which combines the predictions of several trees into a final prediction. In this paper, we use decision trees (torest), AdaBoost and GBM (gradient boosted machine) to build the forest in a stages.

\par More specifically, we use a CART (classification and regression tree) algorithm that builds a set of decision trees by selecting features, in order to divide the training data, by optimizing the entropy in each level. GBM trains a sequence of trees, where each successive tree aims to predict the pseudo-residuals of the preceding trees assuming that the loss function is MSE (mean squared error). This method allows us to combine a large number of regression trees with a small learning rate.  
AdaBoost trains a set of weak learners (decision trees) and combines them into a weighted sum that represents the final outcome. 

\par Since our data is based on samples from 30 subjects, we can utilize the leave-one-out protocol, i.e., the learning process is repeated 30 times, and in each test, 29 subjects are used as a training set and one subject is used as a test set for evaluating the predictive performance of the method. 
\subsection{Detection of Intoxication Using Classification}
The first task in our research is to detect whether a person is intoxicated or not using supervised machine learning algorithms and in order to do this we aim to learn the indicators (of intoxication) from the differences to predict drunkenness. 
We start by using all four devices as a connected network. 
We use movements from the head, hands, and rump obtained by Google Glass, the LG G watch, Microsoft Band, and Samsung Galaxy S4.
The first approach we used was to handle the task of drunkenness detection as a classification task. The output of this task is binary (drunk, sober) and varies depending on the country (based on the BrAC threshold of each country). 

We trained three different models:  GBC (gradient boosting classifier), AdaBoost, and decision trees (DT). We evaluated our models using the leave-one-out protocol against each of the thresholds {220, 240, 250, 350}. 
Table~\ref{tab:table5} shows the AUC score of each model;

\begin{table}
	\centering
	\resizebox{\columnwidth}{!}{%
	\begin{tabular}{cccc}
		{\small\textit{BRAC Threshold}}
		& {\small \textit{AdaBoost}} 
		& {\small \textit{Gradient Boosting Classifier}} 
		& {\small \textit{Decision Trees}} \\
		\midrule
		220 & 0.91 & 0.86 & 0.53  \\\hline
		240 & 0.95 & 0.95 & 0.61  \\\hline
		250 & 0.56 & 0.56 & 0.56  \\\hline
		350 & 0.87 & 0.84 & 0.85  \\\hline
	\end{tabular}%
	}
	\caption{ AUC of classification algorithms (AdaBoost, Decision Trees, and Gradient Boosting Classifier) using all four of the devices in each threshold. }~\label{tab:table5}
\end{table}

The GBC and AdaBoost classifiers yielded excellent results with AUCs of 0.95 at a threshold of 240. AdaBoost also yielded a strong result with an AUC of 0.91 at a threshold of 220.
In order to determine the accuracy of the models, we fixed the threshold at 240.

\begin{figure}
	\includegraphics[width=1.0\columnwidth]{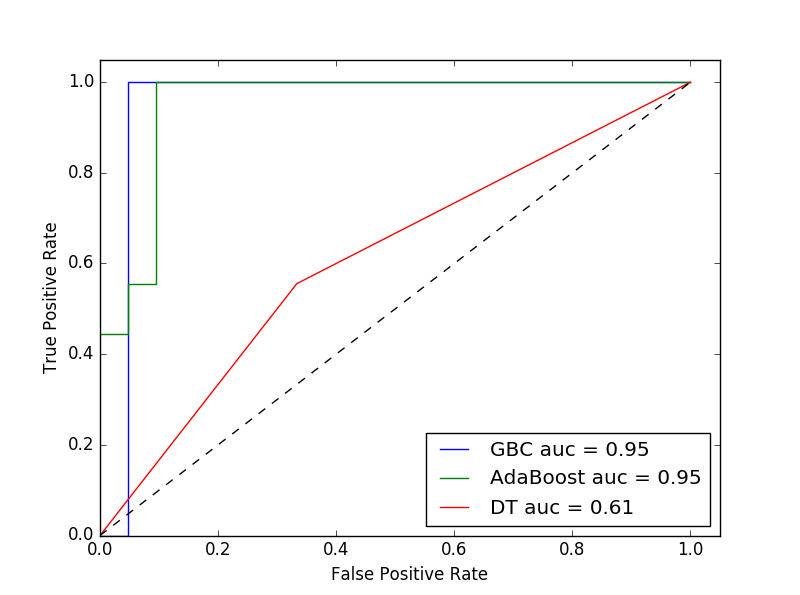}
	\caption{ROC curve of classification algorithms (AdaBoost, Decision Trees, and Gradient Boosting Classifier) using all four of the devices in threshold 240.}~\label{fig:figure66}
\end{figure}

Figure~\ref{fig:figure66} presents the ROC (receiver operating characteristic) curve of the models at the fixed threshold of 240 (using four devices).  The positive class is drunk, and the negative class is sober.

Despite the high AUC that the GBC yields, there is one drunk instance that was misclassified as a sober instance (false negative) as can been seen in the confusion matrix in Table~\ref{tab:table6}. Since this type of mistake can be dangerous (by accidentally allowing a drunk driver to drive), we wanted to verify our results with the limitation that each drunk instance in our data-set will be predicted as drunk. In other words, we fixed the TPR (true positive rate) at 1.0 (the true class is drunk) and checked the damage this limitation caused to the false positive rate (FPR). 
\begin{table}
	\centering
	\begin{tabular}{l c c}
		{\small\textit{}}
		&{\small\textit{Drunk}}
		& {\small \textit{Sober}} \\
		\midrule		
		Predicted as drunk & 8 & 1 \\
		Predicted as sober & 1 & 20 \\
	\end{tabular}
	\caption{Confusion matrix of the gradient boosting classifier with BrAC threshold of 240. }~\label{tab:table6}
\end{table}

Table~\ref{tab:table100} presents the FPR results of the GBC, AdaBoost and Decision Trees classifiers with a fixed TPR of 1.0 at a BrAC threshold of 240.
Since, the GBC yielded the highest results, we fixed it and evaluated this model using the leave-one-out approach against four thresholds: {220, 240, 250, 350} (see Table~\ref{tab:table1}). Table~\ref{tab:table10} shows the FPR of the four thresholds with a fixed TPR of 1.0 using the same data. Since lower values of FPR represent better performance, it can be seen that the GBC yields excellent results, with 100\% detection of intoxication and a minimum number of mistakes (one sober instance that was mistakenly classified as drunk - false positive). This demonstrates that this kind of model can be deployed in order to prevent drunk people from driving with a high rate of intoxication detection (100\% accuracy) and a minimal amount of misclassification (mistakenly classifying sober individuals as drunk).

\begin{table}
	\centering
	\begin{tabular}{lccc}
		&\textit{GBC}
		&  \textit{AdaBoost}
		&  \textit{Decision Trees} \\
		\midrule
		\begin{tabular}[c]{@{}l@{}} FPR  (with \\TPR = 1.0)\end{tabular} & 0.05 & 0.1 & 1.0  \\\hline
	\end{tabular}
	\caption{FPR (false positive rate) of the Gradient Boosting Classifier (GBC),AdaBoost and Decition Trees with fixed TPR (true positive rate) of 1.0 in threshold of 240. }~\label{tab:table100}
\end{table}

\begin{table}
	\centering
	\begin{tabular}{lcccc}

		&\textit{220}
		&  \textit{240}
		&  \textit{250} 
		&  \textit{350} \\
		\midrule
			\begin{tabular}[c]{@{}l@{}} FPR  (with \\TPR = 1.0)\end{tabular} & 0.25 & 0.05 & 0.5 & 1.0 \\\hline
	\end{tabular}
	\caption{FPR (false positive rate) of the Gradient Boosting Classifier with fixed TPR (true positive rate) of 1.0 in each on each drunkenness threshold. }~\label{tab:table10}
\end{table}
 Since most people don't carry smart glass, smart watches, smart fitness trackers, and smart phones all at the same time, we wanted to look at the results based on subsets of the devices.
\par
\begin{figure}
	\includegraphics[width=1.0\columnwidth]{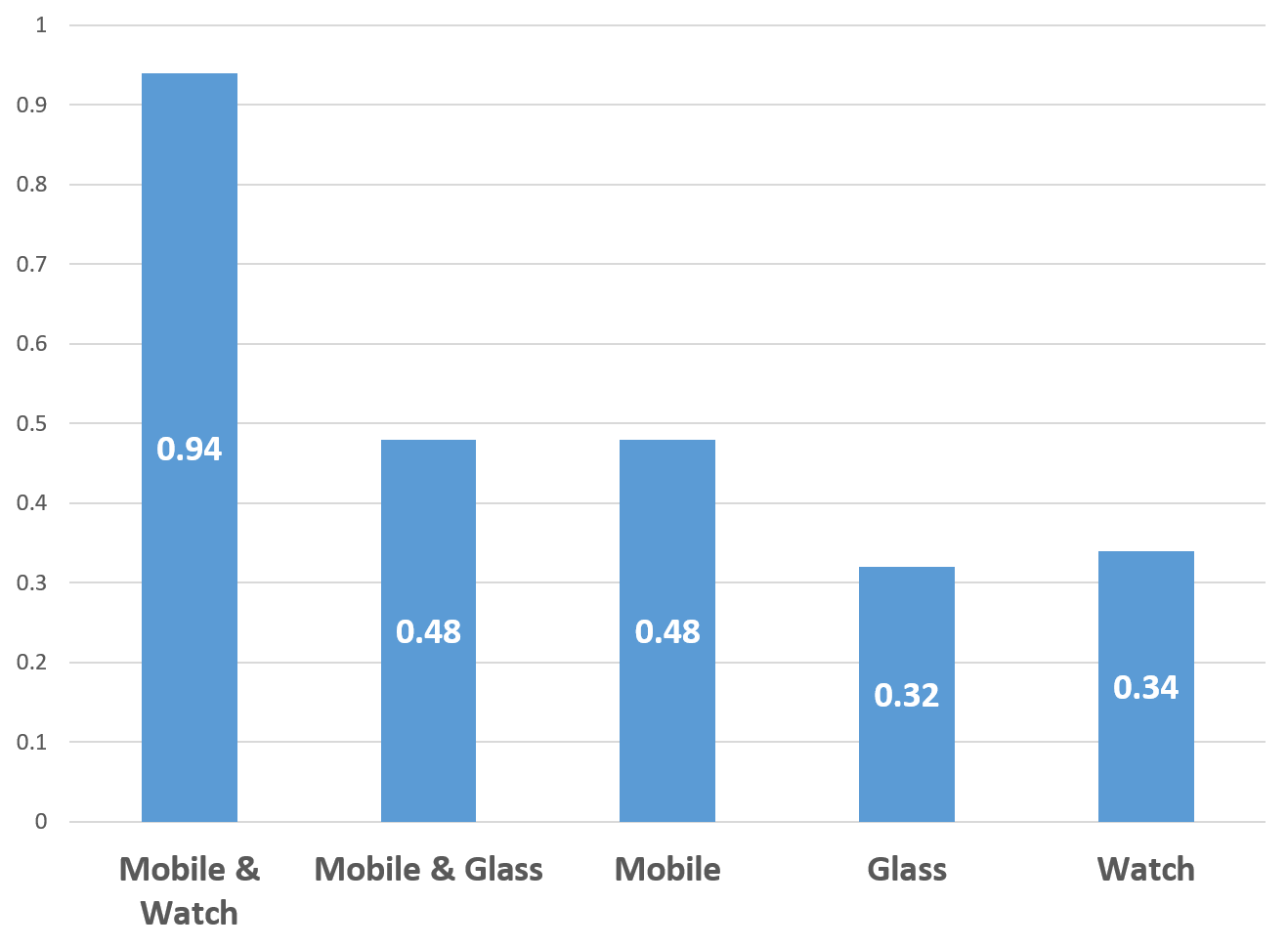}
	\caption{Comparison of AUC for different combinations of devices with GBC and 240 as threshold task.}~\label{fig:figure77}
\end{figure}

 Therefore, we conducted the experiment again, with each device separately (smartphone, glass, smartwatch) and in the following combinations: smartphone/smartwatch and smartphone/Google Glass. 
Figure~\ref{fig:figure77} presents the AUC of each combination at a threshold of 240 using the GBC.
We can see from the results, that the devices when used on their own, are not strong enough to effectively predict drunkenness. However, the results of the smartphone/smartwatch combination (0.94 AUC) were close to those obtained by using all four devices together (0.95 AUC). 
These results have great implications given today's high use of smartphones (80\% usage across the world). Furthermore, a survey conducted in 2014 \cite{GWI-Device-Summary}, indicated that one out of six people around the world owns fitness tracker or smartwatch (with optimistic forecasts for increased use in the future), which means that a model that combines a smartwatch and smartphone could be effectively deployed in the near future.
 
\subsection{Detection of Intoxication Using Regression}
\par The second approach utilized supervised machine learning algorithms for regression to predict the BrAC. The output of this type of task is numeric and can be framed as a virtual breathalyzer based on the gait differences derived from the four devices.
\par We used the same instances from the first approach (the classification task with four devices) and trained four different models, three of which are regression models of the same algorithms used in the classification task: GBR (gradient boosting regression), ABR (AdaBoost regression), and Regression Trees  (RT), as well as Lasso. We evaluated our models using the leave-one-out approach.
\par Table~\ref{tab:table50} presents the MAE and RMSE of each model. This time, the RT resulted in the best performance, as it yielded a MAE of 83 micro-grams of alcohol per one liter of breath (the lowest of all of the models). This result represents a mistake with approximately 19\% of the interval of the data, since the highest label was 430 and the lowest label was 0. 
\begin{table}
	\centering
	\begin{tabular}{lcc}
		 \textit{Classifier}
		&  \textit{MAE}
		&  \textit{RMSE} \\
		\midrule
		Gradient Boosting Regression& 97.2 & 123.3 \\\hline
		Regression Trees  & 83.1 & 124.7 \\\hline
		Lasso & 120.9 & 167.9 \\\hline
		AdaBoost Regression & 103 & 144 \\\hline
	\end{tabular}
	\caption{MAE (mean absolute error) and RMSE (root mean squared error) of different regression algorithms. }~\label{tab:table50}
\end{table}

As was done previously, we wanted to check our ability to predict drunkenness from a subset of devices. We used the same five combinations of devices previously utilized in  the classification approach: (1) smartphone,(2) smartwatch, (3) Google Glass, (4) smartphone/smartwatch combination, and (5) smartphone/Google glass.
\par
\begin{figure}
	\includegraphics[width=1.0\columnwidth]{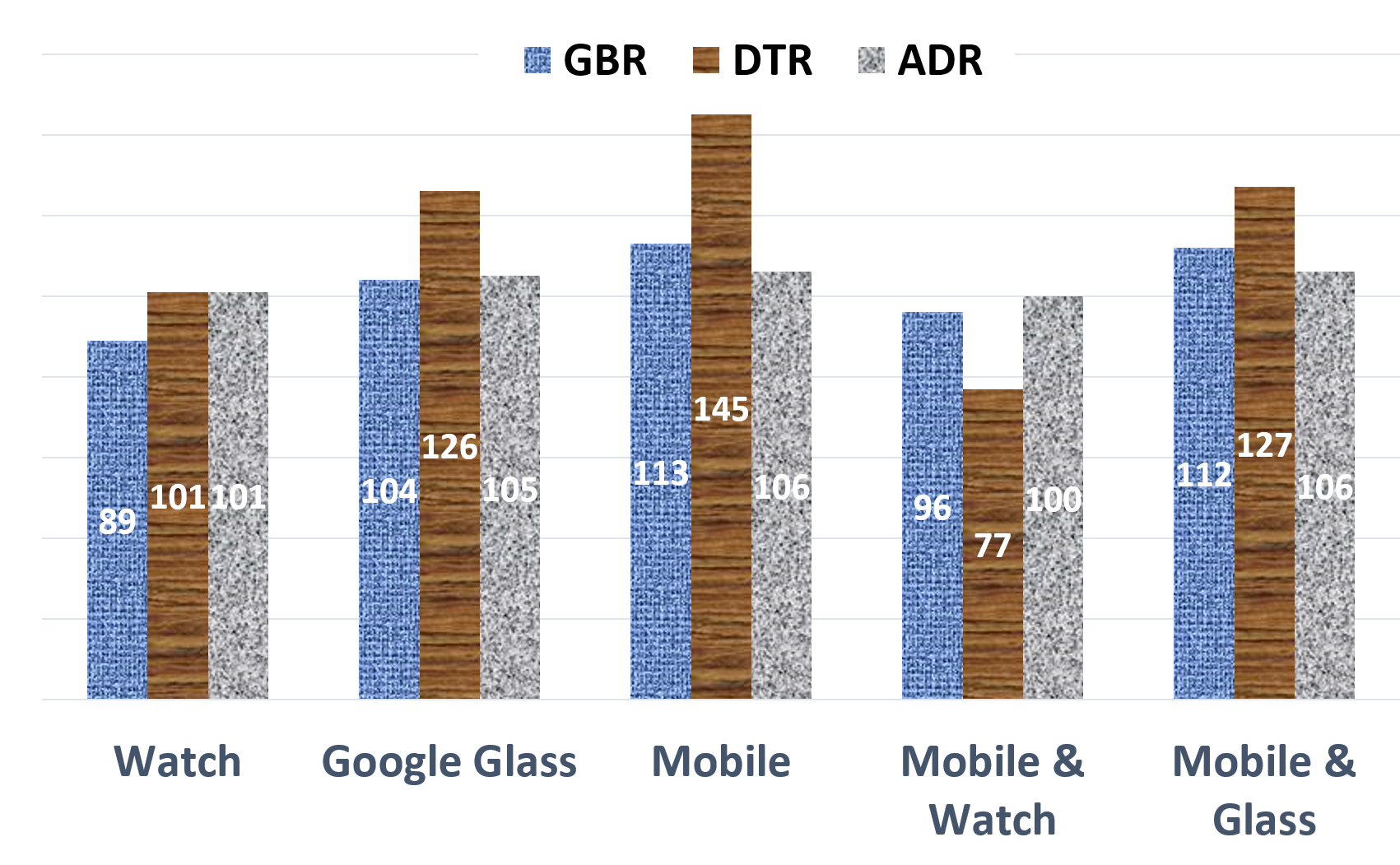}
	\caption{MAE (mean absolute error) of different regression algorithms (GBR, RT, ADR) independently from each device.}~\label{fig:figure99}
\end{figure}

 Again, we evaluated our models using the leave-one-out approach. Figure~\ref{fig:figure99} presents the MAE (mean absolute error) of each model for each device. The combination of the watch and the smartphone also yielded the best performance, as its MAE is minimal across all models.

\subsection{Feature Robustness}
The third task in our research is to detect which set of features yields the highest performance in the task of detection of drunkenness.  In the feature extraction process we extracted four types of features as described in detail in Table~\ref{tab:table3}. Since the gait individuals changes as a result of alcohol consumption, we wanted to identify the best set of indicators to detect drunkenness (based on body movement patterns) and determine whether using 1)the distribution of the movement (histogram), 2) frequency of the movement (FFT), 3) statistics or 4) known gait features, is most effective at this task.
in order to do so, we used the 30 instances with all four devices as we did in the first approach. We evaluated this approach utilizing the classifier that yielded the highest AUC (the GBC) and a threshold of 240. We classified each instance using two methods. The first classification method done using a specific set of features among the sets (histogram, known gait features, frequency features, statistics). The second classification method done using all the other sets of features (except the set used in the first method). Table~\ref{tab:table2000} presents the AUC of each of the sets for the two classification methods.
The histogram is the most important set of features to include, since this set of features yields the highest AUC when used on its own, and the worst AUC is obtained when classification is handled without the histogram set. the Histogram set of features yields the highest AUC when classifying only with it, and the worst AUC when classifying without it, It is the most important set among the other sets. We believe that this is the result of the effects of alcohol consumption on the distribution of movement across the parts of the body.
Since the frequency set of features yielded an AUC score of 0.53, and the highest AUC score obtained when we classified without this set of features, we can only infer that either the frequency of the movement of the body is poor indicator for drunkenness detection, or that we should attempt to detect the frequency of the body as an indicator for drunkenness from another device.

\begin{table}
	\centering
	\begin{tabular}{lcc}
		\textit{Set Of Features}
		&  \textit{Only this set}
		&  \textit{		\begin{tabular}[c]{@{}l@{}}All other sets \\   but this set\end{tabular}} \\
		\midrule
		Histogram Features& 0.80 & 0.42 \\\hline
		Known Gait Features & 0.76 & 0.94 \\\hline
		Frequency Features & 0.53 & 0.97 \\\hline
		Statistics Features & 0.67 & 0.54 \\\hline
	\end{tabular}
	\caption{The AUC scores of different set of features}~\label{tab:table2000}
\end{table}

\section{Discussion}
Excellent results can be achieved when combining various wearable devices for the task of intoxication detection.
Even when combining only smartphone a smartwatch, excellent results are obtained. Since the use of each device on its own does not result in good prediction, we can infer that (1) the differences in the movements of a single body part cannot predict intoxication, and (2) the prediction of intoxication based on movement differences can be achieved with measurements taken from at least two parts of the body
\par We are optimistic that within a few years, with the increased adoption of wearable devices and the ongoing IoT revolution, our system can be implemented on subjects that routinely carry a smartwatch along with their smartphone.
\par Since our system yielded good results predicting intoxication at the thresholds of 220 and 240, it can currently be deployed in countries that utilize this threshold range. 
\section{CONCLUSIONS}
In this paper, we describe a novel approach to detect intoxication from motion differences using wearable devices (smart watch, smart glass, fitness tracker, smartphone). 
We conduct an experiment using 30 people across three different bars in order to evaluate our approach.
\par Supervised machine learning models were trained and result in an AUC of 0.95 for a BrAC threshold of 240 micrograms of alcohol per one liter of breath using four devices and an AUC of 0.94 using only a smart phone and smart watch.
\par A system based on this approach can be used to alert people from driving under the influence of alcohol using two simple gait samples (from a car to a bar and vice-versa) and may also be used to trigger the owner's connected car to prevent ignition in cases in which the owner os detected as drunk.
\par We believe that our results will establish a baseline and also provide an indication of some of the challenges associated with intoxication detection based on free gait. 
\section{Future Work}
There are numerous opportunities to extend this work. One opportunity is to add new devices that are not yet on the market to the system in order to sample more parts of the body such as smart rings and smart shirts.
\par Another opportunity to extend this research is to compare the results of personal models/classifiers versus the collaborative model used in this research. We believe that personal models may result in greater accuracy. In this case, the process can be simplified by generating different models for people with the same height, weight, age, and gender. 

\section{ACKNOWLEDGMENTS}
We would like to thank all of the participants in this study. We also thank A, B, C, and D for their help collecting samples from the subjects.

\balance{}

\bibliographystyle{IEEEtran}
\bibliography{IEEEabrv,sample}
\end{document}